\documentclass[aps,pre,twocolumn,showpacs] {revtex4}
\usepackage{epsfig}
 \usepackage{bm}
\begin{document}
\def\Fbox#1{\vskip1ex\hbox to 8.5cm{\hfil\fboxsep0.3cm\fbox{%
  \parbox{8.0cm}{#1}}\hfil}\vskip1ex\noindent}  
\newcommand{\B}[1]{{\bm{#1}}}
\newcommand{\C}[1]{{\mathcal{#1}}}
\renewcommand{\it}[1]{\textit{#1}}
\newcommand{\Onecol} {\begin{widetext} \onecolumngrid} 
\newcommand{\Twocol} {\end{widetext} \twocolumngrid}   
\title{Conformal Dynamics of Precursors to Fracture}

\author{Felipe Barra$^*$, Mauricio Herrera and Itamar Procaccia}
\affiliation{~Department of Chemical Physics, The Weizmann Institute
of Science, Rehovot 76100, Israel\\ $^*$ Dept. de F\'{\i}sica, Facultad de Ciencias
F\'{\i}sicas y Matem\'aticas, Universidad de Chile, Casilla 487-3, Santiago Chile}
\begin{abstract}
An exact integro-differential equation for the conformal map from
the unit circle to the boundary of an evolving cavity in a
stressed 2-dimensional solid is derived. This equation provides
an accurate description of the dynamics of precursors to fracture
when surface diffusion is important. The solution predicts the
creation of sharp grooves that eventually lead to material failure
via rapid fracture. Solutions of the new equation are
demonstrated for the dynamics of an elliptical cavity and the
stability of a circular cavity under biaxial stress, including
the effects of surface stress.
\end{abstract}
\maketitle

The process of rapid fracture of solids which have failed to
sustain stress is a poorly understood subject \cite{99FM};
elasticity theory does not suffice to describe this process,
since plastic deformations occur at the most interesting
``process zone" where the actual fracture is taking place
\cite{00Lan}. The displacement field is not the only relevant
field, and even if the stress field is given everywhere, it is
not known how the fracture propagates: there exist complex
interactions with sound waves and maybe other fields
\cite{00AKV,01KKL}. On the whole it is not obvious how to achieve
a self consistent theory.

The situation is much clearer when one studies slow processes that
may precede rapid fracture. In particular we will discuss in this
Letter ``precursors" to fracture. Namely, the dynamics of stress
driven deformations of cavities (or free surfaces) in solids
\cite{86Gri,72AT,93YS,94SM,95Gao,97WS,98BM,01BMS,01PS,02XE}. Such
deformations are expected to lead, eventually, to the creation of
deep grooves which then suffer such large stresses that the solid
fails via rapid fracture. The aim of this Letter is to derive and
demonstrate a new equation of motion for the conformal map from
the unit circle to the evolving stress-driven deformed cavity in
2-dimensional solids. This equation offers an accurate
description of the slow evolution of the precursors to failure,
until the moment that rapid fracture can be sustained. In
contrast to all previous treatments we include both surface
energy and surface stress and show that our equation is well
posed.

To set up the problem imagine a 2-dimensional elastic medium with
a hole of an arbitrary shape, whose boundary $z(s)$ is
parametrized by the arc-length variable $s$. Boundary conditions
at infinity load the medium, inducing a displacement field $u(\B
r)$ with a strain  tensor $\epsilon_{jk}$, related to the stress
by \cite{86LL}
\begin{equation}
\sigma_{ij}\equiv \frac{E}{1+\nu}\left(\epsilon_{ij}+
\frac{\nu}{1-2\nu} \delta_{ij}\sum_k\epsilon_{kk}\right) \ .
\label{stress}
\end{equation}
Here $\nu$ is the Poisson ratio and $E$ the Young modulus. Under
stress there begins a process of surface diffusion which deforms
the boundary, with dynamics determined by the velocity $v_n$
normal to the boundary,
\begin{equation}
v_n = -D \frac{\partial^2 \mu}{\partial s^2} \ , \label{vn}
\end{equation}
where $D$ is a diffusion coefficient that depends on the material
and the temperature, and $\mu$ is the chemical potential at the
boundary \cite{57Mul}. The chemical potential is obtained from
the change in total energy associated with an infinitesimal
variation of the interface \cite{72AT,02SB}:
\begin{equation}
\mu = \mu_0
+C[S-\gamma\kappa+\beta\left(\frac{\partial\epsilon_{tt}}{\partial
n}-\kappa \epsilon_{tt}\right)]\ , \label{mu}
\end{equation}
where $\gamma$,~$\kappa$ and $\beta$ are a the surface energy,
the mean curvature and surface stress respectively. $S$ stands
for the deformation energy $S=\sum_{i,j}
\epsilon_{ij}\sigma_{ij}/2$ \cite{94SM}. The notation $t$ and $n$
stands for ``tangent" and ``normal" to the interfce, defined
precisely in Eqs.(\ref{tannor}). $\mu_0$ and $C$ are the reference
chemical potential and a material parameter that we can take as
zero and unity respectively. Note that we differ from
\cite{93YS,94SM,97WS} in taking into account the surface stress.
Ref. \cite{93Gri} incorporated all this physics but performed
stability analysis of the flat interface only.

To evolve the cavity we need to compute then the chemical
potential on its boundary, consistent with the evolving stress
field in the medium. We will assert that on the slow time scale of
surface diffusion the elastic medium is in equilibrium, i.e. $
\sum_k \partial_{k} \; \sigma_{jk} = 0 \quad \mbox{ for all } j $.
The general solution of these equations in 2-dimensional is given
by  \cite{86LL}
\begin{equation}
\sigma_{xx}  =  \partial_y^2 \chi \ , \quad
\sigma_{yy}  =  \partial_x^2 \chi \ , \quad
\sigma_{xy}  =  -\partial_{xy} \chi  \ , \label{sigmachi}
\end{equation}
where the so called Airy potential $\chi$ fulfills the biharmonic
equation $ \Delta^2 \chi = 0$. The general solution of this
equation is written in complex notation, with $z=x+iy$, $\bar z
=x-iy$, as
\begin{equation}
\chi(z,\bar z) = \Re \left[ \overline{z}
\phi(z) + \tilde \psi(z) \right] \ ,\label{solbi}
\end{equation} where
$\phi(z)$ and $\tilde \psi(z)$ are any pair of analytic
functions, to be determined from the boundary conditions.

Consider the boundary of the cavity. With $\alpha$ the angle
between the tangent and the $x$-axis at $z(s)$, define derivatives
in the tangent and normal directions:
\begin{eqnarray}
\partial_t & = & \cos(\alpha) \; \partial_x + \sin(\alpha) \; \partial_y \nonumber \\
\partial_n & = & -\cos(\alpha) \; \partial_y + \sin(\alpha) \; \partial_x \ . \label{tannor}
\end{eqnarray}
The surface stress now must be balanced by the normal component
of the stress \cite{02SB,93Gri}:
\begin{equation}
\partial_{tt} \chi = \sigma_{nn} =\beta \kappa. \quad \mbox{ on the crack} \ . \label{tantan}
\end{equation}
On the other hand the mixed derivatives vanish since there is no restoring force
along the boundary,
\begin{equation}
-\partial_{tn} \chi = \sigma_{tn} = \sigma_{nt} = 0 \quad \mbox{ on the crack} \ . \label{noslip}
\end{equation}
Using these boundary conditions we can evaluate the chemical
potential $\mu$ on the boundary [In principle $\mu$ has terms that cannot be written in terms
of ${\rm Tr}\sigma$ alone, however these terms add to zero as a consequence of 
$\partial_n \sigma_{nn}=-\partial_t \sigma_{nt}$ and $d\sigma_{nt}/ds=0$ on the boundary]
\begin{equation}
\mu=\left[ \frac{1-\nu^2}{E}\left(\frac{({\rm Tr}
\sigma)^2}{2}+\beta \partial_n {\rm Tr} \sigma- \beta \kappa {\rm
Tr} \sigma\right)-\gamma \kappa \right] \label{chem-pot2}
\end{equation}
Using the fact that $4\partial^2\chi/\partial z\partial\bar
z=\sigma_{xx}+\sigma_{yy}= \sigma_{tt}+\sigma_{nn}$ we can
immediately read from Eq. (\ref{solbi}),
\begin{equation}
\text{Tr}\sigma=4\Re[\phi'(z)] \ . \label{solution}
\end{equation}
Thus to compute $\mu$ and its derivatives and advance the cavity
we only need to determine the function $\phi(z)$. The boundary
conditions (\ref{tantan}) and (\ref{noslip}) are expressed in
terms of $\phi(z)$ and $\tilde \psi(z)$ by using Eqs.
(\ref{tannor}) -(\ref{noslip}). We note that
$(\cos\alpha+i\sin\alpha)(\partial_t-i\partial_n)=2\partial_{\bar
z}$ and on the boundary
\begin{equation}
\partial_t\partial_{\bar z}\chi = \frac{1}{2}(\cos\alpha+i\sin\alpha)\partial_{tt}\chi=
\frac{\beta\kappa}{2} \partial_t z \ , \label{bc}
\end{equation}
where we identify $\partial_t z(s)$ as the unit vector tangent to
the boundary. Writing the mean curvature as $\kappa =
\partial_t^2 z(s)/i\partial_t z(s)$, this condition reads
$\partial_t \{\partial_{\bar z} \chi +i\beta \partial_t
z(s)/2\}=0$. Thus the boundary condition on the interface
\cite{52Mus} is
\begin{equation}
\phi(z(s)) + z(s) \; \overline{\phi^\prime(z(s))} +
\overline{\psi(z(s))} = -i\beta \partial_t z(s) + K \ , \label{eq}
\end{equation}
where $\psi(z)\equiv \tilde \psi'(z)$ and $K$ is a constant (that can be chosen
zero with impunity).

For the boundary conditions at infinity we consider biaxial
loading:
\begin{equation}
\sigma_{xx}(\infty)=\sigma_0\ , \quad
\sigma_{yy}(\infty)=\sigma_0\ , \quad \sigma_{xy}(\infty)=0 \ .
\label{b.c.inf}
\end{equation}
It was shown in \cite{52Mus} that if the integral over the boundary of the RHS of Eq. (\ref{eq})
is zero, then
the finiteness of the stresses at infinity implies that the analytic functions have a Laurent
expansion of the form
\begin{equation}
\begin{array}{c}
\phi(z) = \phi_1 z +\sum_{i=0}\phi_{-i}z^{-i} \\
\psi(z) = \psi_1 z +\sum_{i=0}\psi_{-i}z^{-i}
\end{array}
\end{equation}
The freedoms that we have in determining the Airy function
(\ref{solbi}) allow us to choose $\phi_0=0$ and $\phi_1$ real.
Then using the boundary conditions (\ref{b.c.inf}) at infinity,
we find $\phi_1=\sigma_0/2$, $\psi_1=0$.
\begin{figure}
\centering
\includegraphics[width=.35\textwidth]{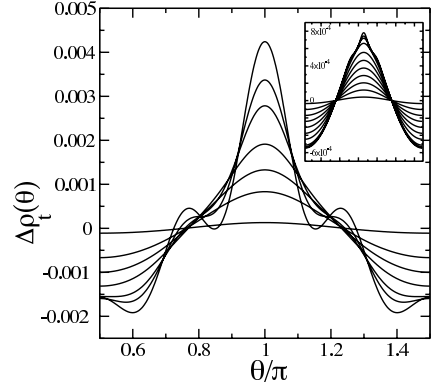}
\caption{Evolution of an ellipse under a biaxial load. Shown is
the change in the radius $\Delta\rho_t(\theta) \equiv
|\Phi(\epsilon,t)|-|\Phi(\epsilon,0)|$ (in dimensionless units)
relative to its initial value. The plot is given at dimensionless
times $10^4\times t=4, 30, 40, 58.9,71.8. 77.0$ and $82.3$. In
the insert we show the analogous evolution for a cavity without
surface stress.} \label{evolution}
\end{figure}
To proceed consider now a conformal map $\Phi(\omega,t)$ which maps the
exterior of the unit circle $\epsilon=\exp(i\theta)$ to the exterior of
boundary $z(s)$.
Our central aim in this
Letter is to derive an equation of motion of this map. The normal velocity $v_n$ is related to the
conformal map as follows: denote by $\hat n=n_x+in_y$ and $z(s)=x(s)+iy(s)$ the unit vector normal to the
boundary, and the position of the boundary respectively . Then
\begin{equation}
v_n(s)\! =\! \dot x(s) n_x+\dot y(s) n_y \!=\!\Re\left[\frac{d z(s)}{dt} \bar{\hat n}\right]\!=\!
\Re\left[\frac{d \Phi(\epsilon,t)}{dt} \bar{\hat n}\right] \ . \label{vnconf}
\end{equation}
(Note that below we use $\epsilon$ and $\exp(i\theta)$ interchangeably).
The tangent ($\hat t$) and normal unit vector are given in terms of the conformal
map by
\begin{eqnarray}
\hat t&=&\frac{\partial \Phi(\epsilon)}{\partial s}=\frac{d\theta}{ds}
\frac{\partial \Phi(\epsilon)}{\partial \theta}=\frac{1}{|\Phi ' (\epsilon)|}\frac{\partial
\Phi(\epsilon)}{\partial \theta}\nonumber\\
\hat n &=& -i\hat t  \ . \label{tn}
\end{eqnarray}
Using this in Eq.(\ref{vnconf}) and remembering that $d\Phi(e^{i\theta})/dt =\partial_t
\Phi + \Phi ' e^{i\theta} i \theta_t$ we derive:
\begin{equation}
v_n(s) = \Re \left( \partial_t
\Phi(\epsilon) \overline{\epsilon \frac{\Phi ' (\epsilon)}{|\Phi ' (\epsilon)|}}
\right) \ .
\end{equation}
We rewrite this equation,
\begin{equation}
\label{temp}
\partial_t \Phi(\epsilon) \overline{\epsilon \frac{\Phi ' (\epsilon)}{|\Phi ' (\epsilon)|}}  =v_n+iC \ ,
\end{equation}
with an unknown imaginary part $C$. Multiplying the last equation by $\epsilon\Phi'/|\Phi'|$ we get the equivalent
equation
\begin{equation}
\partial_t \Phi = \epsilon \Phi'(\epsilon) \left( \frac{v_n}{|\Phi'|}+iC'\right)
\end{equation}
and unknown $C'$. This equation, valid on the interface, can be analytically continued outside the unit circle.
We need to
choose $C'$ such that the term in the parenthesis is an analytic function, removing the freedom in $C'$.  This
also fixes
the parameterization which was so far arbitrary. All this is achieved with the Poisson integral formula,
\begin{equation}
\partial_t \Phi = \omega \Phi'(\omega) \int_0^{2\pi}\frac{d\theta}{2\pi}
\frac{\omega+e^{i\theta}}{\omega-e^{i\theta}}\frac{v_n(e^{i\theta})}{|\Phi'(e^{i\theta})|}
\end{equation}
The equation, being analytic, must have analytic solutions which
provide the dynamics of the conformal map. In practice, the
Poisson integral formula is best expressed in terms of the
Fourier components of the function. Given a real function on the
unit circle
\begin{equation}
g=\frac{v_n(e^{i\theta})}{|\Phi'(e^{i\theta})|}=a_0+\sum_{n \geq 1} (a_ne^{in\theta}+{\bar a}_n e^{-in\theta}) \ ,
\label{eqg}
\end{equation}
the analytic function outside the circle whose real part on the
unit circle is $g$ is
\begin{equation}
G=  \int_0^{2\pi}\frac{d\theta}{2\pi}
\frac{\omega+e^{i\theta}}{\omega-e^{i\theta}}g(e^{i\theta})=a_0+2\sum_{n\geq 1}\bar a_n \omega^{-n}
\label{eqG}
\end{equation}
Thus finally the equation of the conformal map reads
\Fbox{
\begin{equation}
\partial_t \Phi = \omega \Phi'(\omega) G(\omega) \ . \label{final}
\end{equation}}
The conformal map itself (which is univalent) has a Laurent
expansion of the form
\begin{equation}
\Phi(\omega,t) = F_1(t) ~\omega + F_0(t) +\sum_{n=1}^\infty F_{-n}(t) ~\omega^{-n} \ . \label{Laurent}
\end{equation}
By taking derivatives with respect to $t$ and $\omega$ and substituting back in Eq.(\ref{final})
we reduce the dynamics to an infinite set of ordinary differential equations for the
Laurent coefficients $F_i(t)$.  All that remains is to compute the function $g(\epsilon)$
that is given in terms of $v_n(\epsilon)$ and the conformal map.
\begin{figure}
\centering
\includegraphics[width=.35\textwidth]{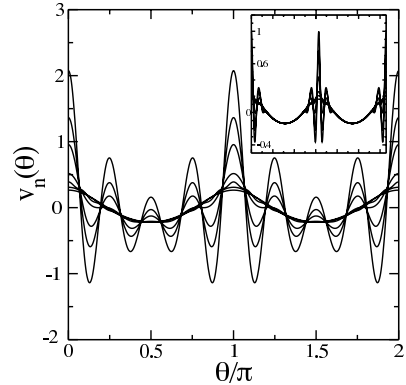}
\caption{The normal velocity as a function of the angle $\theta$
on the unit circle. The times are the same as in Fig.
\ref{evolution}. The insert shows the analogous figure for a
cavity without surface stress. } \label{velocity}
\end{figure}
To compute the normal velocity $v_n(\epsilon)$ first compute ${\rm
Tr} \sigma$ on the boundary. Rewrite Eq.(\ref{eq}) in terms of the
conformal map at time $t$:
\begin{eqnarray}
\phi(\Phi(\epsilon,t)) &+& \frac{\Phi(\epsilon,t)}{\overline{\Phi^\prime(\epsilon,t)}}
\overline{\phi^\prime(\Phi(\epsilon,t))} + \overline{\psi(\Phi(\epsilon,t))} \nonumber\\
&=& \beta
\epsilon\frac{\Phi^\prime(\epsilon,t)}{|\Phi^\prime(\epsilon,t)|}
\ . \label{bcphi}
\end{eqnarray}
According to Eq.(\ref{Laurent}) the functions in Eq. (\ref{bcphi}) read
\begin{eqnarray}
\phi(\Phi(\epsilon,t)) &=& \phi_1 F_1(t)\epsilon +\phi_0(\epsilon)\nonumber\\
\psi(\Phi(\epsilon,t)) &=& \psi_1 F_1(t)\epsilon +\psi_0(\epsilon)\ , \label{phi0}
\end{eqnarray}
where the only unknowns now are the expansion coefficients of the functions $\phi_0$
and $\psi_0$,
\begin{equation}
\phi_0(\epsilon) = \sum_{n=0}^\infty u_{-n} \epsilon^{-n}\ ; \quad
\psi_0(\epsilon) = \sum_{n=0}^\infty v_{-n} \epsilon^{-n} \ .
\end{equation}
At this point we need to compute the two Fourier transforms
\begin{equation}
\frac{\Phi(\epsilon,t)}{\overline{\Phi^\prime(\epsilon,t)}} =\sum_{n=-\infty}^\infty b_j \epsilon^j \ ; \quad
\frac{\Phi^\prime(\epsilon,t)}{|\Phi^\prime(\epsilon,t)|} =\sum_{n=-\infty}^\infty c_j \epsilon^j \ ,
\end{equation}
substitute them in to Eq. (\ref{bcphi}), and obtain a system of equations for the coefficients
$u_{-n}$ and $v_{-n}$. In fact upon the substitution one finds that the determination of the function
$\phi(\omega)$ is independent of $\psi(\omega)$, requiring only negative powers of $\epsilon$.

Having found the function $\phi(\omega)$ we employ Eq.
(\ref{solution}) to compute the trace of the stress tensor
anywhere on the boundary. The curvature $\kappa$ is:
\begin{equation}
\kappa=\Re\left(\frac{1}{|\Phi'|}(1+\frac{\Phi''}{\Phi'}e^{i\theta})\right) \ . \label{kappa}
\end{equation}
Taking the second derivative of the chemical potential we have $v_n(\epsilon)$. With this
function at hand we can return to Eq. (\ref{eqg}) and step the equations for the
Laurent coefficients, finding the new conformal map, etc.
\begin{figure}
\centering
\includegraphics[width=.35\textwidth]{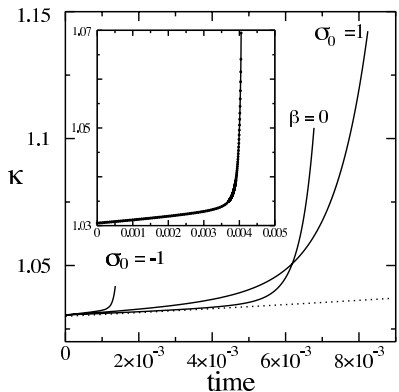}
\caption{The curvature at the tip as a function of time for an
ellipse under compression and extension, and for zero surface
stress. Dotted line: the curvature predicted by the linear
stability analysis.} \label{curvature}
\end{figure}

To demonstrate the efficiency of this procedure, and the
interesting predictions it offers for the dynamics of cavities in
stressed solids, we present (i) the evolution of an initial
elliptical hole and (ii) the stability analysis of a circular
hole. For (i) our initial conformal map is
\begin{equation}
\Phi(\omega, t=0) = F_1(0)\omega+\frac{F_{-1}(0)}{\omega} \ ,
\end{equation}
where we chose $ F_1(0)=1$, $F_{-1}(0)=0.01$. For material
parameters we took $(1-\nu^2)/E=1$, $\gamma=0.2$, $D=1$ and
$\beta=0.1$, all in dimensionless units. We study the evolution
of the initial ellipse under biaxial load with $\sigma_0=1$. In
Fig. \ref{evolution} we show the time evolution of $\rho_t\equiv
|\Phi(\epsilon,t)|-|\Phi(\epsilon,0)|$. In the inset we show the
analogous dynamics when $\beta=0$, $\gamma=0.1$. We see that the
tip of the ellipse is advancing at the expense of two dips that
develop on its side, but this effect is more dramatic for
$\beta=0$ as seen in the inset. In Fig.2 we show the normal
velocity, with the inset for the same parameters as in Fig.1.
Finally, in Fig. \ref{curvature} we show the curvature at the
tip. When the stress at the tip reached the Griffith criterion
the material would yield by rapid fracture \cite{93CG}.

For (ii) we start with
\begin{equation}
\Phi(\omega,t)=R\omega+\sum_{n=1}f_n(t)\omega^{-n} ;\quad |f_n|<<1
\quad \forall n\ ,
\end{equation}
which maps the unit circle ($\epsilon=e^{i\theta}$) to a wavy
shaped circle of radius $R$. Using our equation (\ref{final} for
the conformal map, and Eq. (\ref{bcphi}) which determines the
stress, linearizing in small $f_n$, it is a straightforward
calculation to obtain the stability eigenvalues in the form
\begin{widetext}
\begin{equation}
\lambda_n=\frac{(n+1)^2}{R^2}\Bigg\{\frac{1-\nu^2}{E}
\left[\left(\frac{8\sigma_0^2}{R}-\frac{12\beta\sigma_0}{R^2}
\right)n-\left(\frac{4\beta^2}{R^3}+\frac{2\beta\sigma_0}{R^2}
\right)n^2-2\frac{\beta^2}{R^3}n^3
\right]-\frac{2\gamma}{R^2}n-\frac{\gamma}{R^2}n^2\Bigg\} \ .
\label{lambdan}
\end{equation}
\end{widetext}
Contrary to the calculation in \cite{93Gri} our problem is well
posed, the cubic term ($n^3$) is negative. The analytic result
(\ref{lambdan}) is novel, showing precisely what load is needed
to destabilze  a circular cavity.

In summary, we stress that surface stress term in $\mu$ may
matter: it introduces for sharp tips a competitive term
proportional to $r^{-3/2}$ in contrast to the terms proportional
to $1/r$ of the stress energy and the curvature. surface stress
removes the degeneracy of compression and extension, as we see in
Eq. (\ref{lambdan}). In the nonlinear regime the grooves start to
form faster for compression. These and related issues will be
elaborated further in a future publication.

 We thank Efim Brener for suggesting this problem to
us, and for some useful e-discussions with him and with R.
Spatschek. This work was supported in part by the Minerva
Foundation, the European Commission under a TMR grant, and the
Fundaci\'on Andes under project c-13760.

\end{document}